\documentclass[11pt,epsf]{article}

\usepackage{epsfig}
\usepackage{amssymb}
\usepackage{graphicx}
\usepackage{color}
\usepackage{subfigure}
\usepackage{hyperref}

\def\be{\begin{equation}}\def\ba{\begin{eqnarray}}
\def\ee{\end{equation}}\def\ea{\end{eqnarray}}

\makeatletter

\@addtoreset{equation}{section} \makeatother
\def\be{\begin{equation}}
\def\ee{\end{equation}}
\def\ba{\begin{array}}
\def\ea{\end{array}}
\def\bea{\begin{eqnarray}}
\def\eea{\end{eqnarray}}
\def\nn{\nonumber\\}

\def\ct{\cite}
\def\la{\label}
\def\eq#1{(\ref{#1})}

\def\a{\alpha}
\def\b{\beta}

\def\d{\delta}
\def\D{\Delta}
\def\ep{\epsilon}

\def\ph{\phi}
\def\Ph{\Phi}
\def\ps{\psi}

\def\k{\kappa}

\def\m{\mu}

\def\s{\sigma}

\def\o{\omega}
\def\O{\Omega}

\def\half{\frac{1}{2}}

\def\pa{\partial}

\def\lb{\left[}
\def\lc{\left\{}
\def\ls{\left(}

\def\rb{\right]}
\def\rc{\right\}}
\def\rs{\right)}

\def\td#1{\tilde{#1}}

\def\sss{\scriptscriptstyle}

\begin{document}
\begin{titlepage}
\title{\vskip -60pt
\vskip 20pt Soft Wall Model in the Hadronic Medium }
\author{Chanyong
Park$^a$\footnote{e-mail : cyong21@sogang.ac.kr}, Do-Young
Gwak$^b$\footnote{e-mail : dyGwak@sogang.ac.kr }, Bum-Hoon
Lee$^{ab}$\footnote{e-mail : bhl@sogang.ac.kr}, Yumi
Ko$^c$\footnote{e-mail : koyumi@apctp.org}, and Sunyoung
Shin$^d$\footnote{e-mail : shin@theor.jinr.ru}}
\date{}
\maketitle \vspace{-1.0cm}
\begin{center}
~~~
\it $^a\,$Center for Quantum Spacetime, Sogang University, Seoul 121-742, Korea\\
\it $^b\,$ Department of Physics, Sogang University, Seoul 121-742, Korea \\
\it $^c\,$Asia Pacific Center for Theoretical Physics, Pohang, Gyeongbuk 790-784, Korea \\
\it $^d\,$Bogoliubov Laboratory of Theoretical Physics, JINR, 141980
Dubna, Moscow region, Russia
~~~\\
~~~\\
\end{center}

\begin{abstract}
We study the holographic QCD in the hadronic medium by using the
soft wall model. We discuss the Hawking-Page transition between
Reissner-Nordstr\"{o}m AdS black hole and thermal charged AdS of
which the geometries correspond to deconfinement and confinement
phases respectively. We also present the numerical result of the
vector and axial vector meson spectra depending on the quark
density.
 \end{abstract}

{\small
\begin{flushleft}
\end{flushleft}}
\end{titlepage}
\newpage

\tableofcontents 
\section{Introduction}

Holographic QCD offers new approaches to understand the strongly
interacting regime of gauge theories based on AdS/CFT duality
\cite{Maldacena:1997re}. The properties of QCD including confinement
and chiral symmetry breaking have been discussed in asymptotically
AdS geometries
\cite{Witten:1998zw,Polchinski:2000uf,Klebanov:2000hb,Maldacena:2000yy,
Polchinski:2001tt,Kruczenski:2003uq,Sakai:2004cn}.

The QCD-like models can be derived in both top-down and bottom-up
approaches. There are two methods to construct AdS/QCD background in
the bottom-up approach. One is introducing an infrared (IR) cutoff
\cite{Polchinski:2001tt,BoschiFilho:2002vd,deTeramond:2005su,Erlich:2005qh,Da
Rold:2005zs,Ko:2009jc,Lee:2010dh}, and the other is scaling the
action or modifying the background metric by introducing a scalar
field or a warp factor so that a smooth IR truncation is induced
\cite{Karch:2006pv,Andreev:2006vy,Forkel:2007cm}. The light-front
holography \cite{Brodsky:2003px} is constructed by using the both
methods.

The phase transition between the confining and deconfining phases
corresponds to the Hawking-Page transition \cite{Hawking:1982dh} of
gravity between thermal AdS (tAdS) at low temperature and the
Schwarzschild AdS black hole at high temperature in pure gauge
theory \cite{Witten:1998zw}. The phase transition of holographic QCD
in the hard wall \cite{Erlich:2005qh} and soft wall model
\cite{Karch:2006pv} was discussed in \cite{Herzog:2006ra}.

The holographic QCD can be improved by including the chemical
potential
\cite{Kim:2006gp,Horigome:2006xu,Nakamura:2006xk,Kobayashi:2006sb,
Domokos:2007kt,Kim:2007em,Sin:2007ze,Lee:2009bya,Park:2009nb,Jo:2009xr,Seo:2009kg,
Andreev:2010bv,Bigazzi:2011it}.

The charge of Reissner-Nordstr\"{o}m AdS black hole (RNAdS BH) can
be interpreted as quark charge when branes fill the bulk
\cite{Sin:2007ze}. The boundary value of the time-component gauge
field is the chemical potential whose dual operator is the quark
number density. The gravity dual to deconfining phase of QCD is the
RNAdS BH. The thermal charged AdS (tcAdS) space, which is the zero
mass limit of the RNAdS BH, was proposed as the dual geometry
corresponding to the confining phase of QCD \cite{Lee:2009bya}.
\footnote{By generalizing the flavor gauge symmetry to $U(N_f)_L
\times U(N_f)_R$ and adding a Chern-Simons term, the anomaly of U(1)
axial symmetry \cite{Witten:1998qj} was described in
\cite{Domokos:2007kt}, in which the baryon number is identified with
the instanton number defined from the Chern-Simons term. In this
paper, we will not consider this Chern-Simons term which may affect
nontrivially the dual geometry.} The Hawking-Page transition between
confinement and deconfinement phases was studied in the presence of
the chemical potential or the quark density in the hard wall model
\cite{Lee:2009bya,Park:2009nb}. It has been also observed that meson
mass increases as the quark density increases in the hard wall model
\cite{Jo:2009xr}.

In this report, we first discuss the holographic QCD in the hadronic
(or quark) medium by using the soft wall model \cite{Karch:2006pv}
where a non-dynamical scalar field is introduced to give a smooth IR
truncation. The scalar field does not affect the dynamics of the
gravity as it scales the full action. We next investigate
numerically the spectra of the vector and axial-vector mesons
depending on the quark density, where the correct boundary
conditions consistent with the analytic result of the vector meson
are used.

The rest of this paper is organized as follows. In section 2, we
investigate the Hawking-Page transition between RNAdS BH and tcAdS.
In section 3, we study the vector and axial vector meson spectra
depending on the quark density. In section 4, we summarize our
results and discuss issues for future work.

\section{Hawking-Page transition in the soft wall model \la{sec:hp}}

We consider the holographic QCD in the hadronic (or quark) medium
using the soft wall model. According to the AdS/CFT correspondence,
the quark number operator in the gauge theory side corresponds to
the time-component of the bulk gauge field. So the dual geometry of
the hadronic medium should be a five-dimensional spacetime
containing some electric charges. There are well-known examples
describing the asymptotic AdS space containing the electric charges.
One is the Reissner-Nordstr\"{o}m AdS black hole (RNAdS BH)
corresponding to the deconfining phase described by quarks and
gluons in the gauge theory side. The RNAdS BH in Euclidean spacetime
signature is \bea \la{metric:RNAdSBH} ds^2 = \frac{R^2}{z^2} \ls
f_{RN}(z) dt_{\sss{E}}^2 + \frac{1}{f_{RN}(z)} dz^2 + \d_{ij} dx^i
dx^j \rs,\eea with \be \la{f:RN} f_{RN} (z) = 1 - m z^4 + q^2 z^6,
\ee where $i,j$ imply the
three-dimensional spatial indices. The parameters $R$, $m$ and $q$
are the AdS radius, the black hole mass and charge, respectively. We follow the convention of \cite{Park:2009nb}.
The time-component of the corresponding bulk gauge field is \be
\la{sol:gauge} A_0 = i \ls 2 \pi^2 \m - Q z^2 \rs , \ee where $\m$
and $Q$ are related to the chemical potential and quark number
density in the dual gauge theory. To satisfy the Einstein and
Maxwell equations the quark number density $Q$ should be related to
the black hole charge $q$ \be Q = \sqrt{\frac{3 g^2 R^2}{2 \k^2}} q
. \ee

In the confining phase the vacuum of the dual gauge theory is filled
up with hadronic matter as quarks can not exist alone due to the
confinement. The dual geometry of it was found in
\cite{Lee:2009bya}, which is another solution satisfying Einstein
and Maxwell equations. It is called a thermal charged AdS (tcAdS)
space \cite{Lee:2009bya}, whose metric in Euclidean signature is \be
\la{metric:tcAdS} ds^2 = \frac{R^2}{z^2} \ls f_{tc}(z)
dt_{\sss{E}}^2 + \frac{1}{f_{tc}(z)} dz^2 + \d_{ij} dx^i dx^j \rs,
\ee with \be \la{met:fact} f_{tc} (z) = 1 + q^{\prime 2} z^6 . \ee
The time-component of the bulk gauge field is also in the form of
\eq{sol:gauge}. As will be shown, although this tcAdS solution is
singular at $z = \infty$ it does not give any problem to calculate
the physical quantities due to the potential barrier caused by the
non-dynamical scalar field in the soft wall model.

The Hawking-Page transition in the hadronic medium by using the hard
wall model has been investigated \cite{Lee:2009bya,Park:2009nb}. The
Hawking-Page transition in the soft wall model is also studied for
the case of $q=0$ \cite{Karch:2006pv,Herzog:2006ra}. We investigate
the Hawking-Page transition between the RNAdS BH and tcAdS
backgrounds in the soft wall model, which describes the
deconfinement phase transition of QCD in the hadronic medium. On
these backgrounds, the Euclidean gravity action of the soft wall
model is given by
\begin{equation}    \la{act:Euclidean}
S = \int d^5 x \sqrt{G} e^{-\Phi}\mathrm{Tr} \left[ \,\frac{1}{2
\kappa^2} \,(\,- {\cal{R}} + 2 \Lambda \,) + \frac{1}{4 g^2} F_{MN}
F^{MN}\,\right],
\end{equation}
where the non-dynamical scalar field is $\Phi = c z^2$. According to
the AdS/CFT correspondence, the on-shell gravity action corresponds
to the thermodynamic energy of the dual gauge theory, so we evaluate
the on-shell gravity action on \eq{metric:RNAdSBH} and
\eq{metric:tcAdS} with the gauge field \eq{sol:gauge}.

For the deconfining phase, which is dual to the RNAdS BH, we impose
the Dirichlet boundary condition $A_0 = i 2 \pi^2 \m$ at the
boundary $z=0$ to get the corresponding on-shell gravity action
\begin{eqnarray}
S_{RN} =\frac{R^3 V_3}{\kappa^2} \int_0^{\b_{RN}} dt_{\sss{E}}
\int_{\ep}^{z_+} dz\, e^{- c z^2} \left( \frac{4}{z^5} - 2 q^2
z\right),
\end{eqnarray}
where $V_3$, $\ep$ and $\b_{RN}$ are the volume of the
three-dimensional space, the UV cut-off and the inverse Hawking
temperature $\b_{RN} = 1/T_{RN}$.

As shown in \cite{Lee:2009bya,Park:2009nb}, this on-shell gravity
action is proportional to the grand potential of the dual gauge
theory, which is a function of $\m$. The black hole charge should
therefore be a function of the chemical potential $\m$. To determine
this relation, we impose the regularity condition, $A_0=0$, at the
black hole horizon. Then, the black hole charge or the quark number
density can be rewritten as \be \la{rel:qmu} q = \sqrt{\frac{2
\k^2}{3 g^2 R^2}} \ \frac{2 \pi^2 \m}{z_+^2} , \ee where $z_+$ means
the black hole horizon. The on-shell gravity action becomes \be
\la{act:rnd} S_{RN} =\frac{R^3 V_3}{\kappa^2} \frac{1}{T_{RN}}
\left( F(z_+) - F(\epsilon) + H(z_+) - H(\epsilon) \frac{}{} \right)
, \ee where
\bea%
F(z) &\equiv&  \int  dz\,
 \frac{4\,e^{- c z^2}}{z^5} =\frac{e^{-c z^2}}{z^4} \,(c z^2 -1) + c^2 {\rm Ei}(-c z^2), \nonumber\\
H(z) &\equiv&  - 2 q^2 \int dz \  z \ e^{- c z^2}  = \frac{e^{-c
z^2} q^2}{c} ,
\eea%
with the exponential integral function ${\rm Ei} (z)$ \be {\rm Ei}
(z) = - \int_{-z}^{\infty} dt \  \frac{e^{- t}}{t} . \ee In
\eq{act:rnd}, $F(\ep)$ diverges in the limit as $\ep \to 0$, whereas
the other terms remain finite. We renormalize the on-shell gravity
action \eq{act:rnd} to have a well-defined grand potential of the
boundary theory. We subtract the action of thermal AdS (tAdS or
Euclidean AdS space) corresponding to the reference geometry, whose
on-shell action is
\bea%
\la{act:AdS} S_{AdS} &=& \frac{R^3 V_3}{\kappa^2} \int_0^{\beta_0}
dt_{\sss{E}} \int_{\epsilon}^{\infty} dz\, \frac{4\,e^{- c
z^2}}{z^5} \nn &=& - \frac{R^3 V_3}{\kappa^2} \beta_0  F(\epsilon) ,
\eea%
where $\b_0$ is the time periodicity in tAdS. $F(\infty)=0$ has been
used. Matching the circumferences between the RNAdS BH and tAdS at
the UV cut-off $\ep$, $\b_0$ can be rewritten as
\bea%
\b_0 = \frac{\sqrt{f_{RN} (\epsilon)}}{T_{RN}}
= \frac{1}{T_{RN}} \ls 1 - \frac{m}{2} \ep^4 \rs + {\cal O} (\ep^6).
\eea%
Using the expansion form of $F(\ep)$
\be%
F (\ep) = - \frac{1}{\ep^4} + {\cal O} \ls \frac{1}{\ep^2} \rs ,
\ee%
the renormalized on-shell action ${\bar  S}_{RN}$ in the $\ep \to 0$
limit becomes
\bea%
{\bar  S}_{RN} &=& S_{RN}- S_{AdS} \nonumber\\
&=& \frac{R^3 V_3}{\kappa^2} \frac{1}{T_{RN}} \left[ F(z_+) +
\frac{m}{2} + H(z_+) - \frac{q^2}{c} \right]\la{act:RNren} .
\eea%
As \eq{f:RN} satisfies $f(z_+)=0$, we get the black hole mass $m$ as
a function of the black hole horizon $z_+$
\bea %
m &=& \frac{1}{z_+^4} + q^2 z_+^2 .\la{m:RN}
\eea%
Furthermore, the black hole horizon $z_+$ can be expressed in terms
of the Hawking temperature and chemical potential as follows
\be %
\la{rel:zp} z_+ = \frac{3 g^2 R^2}{8 \pi^4 \k^2 \m^2} \ls
\sqrt{\pi^2 T_{RN}^2 + \frac{16 \pi^4 \k^2 \m^2}{3 g^2 R^2}} - \pi
T_{RN}\rs .%
\ee %
Using the relations \eq{rel:qmu} and \eq{m:RN}, the renormalized
on-shell action \eq{act:RNren} in terms of $z_+$ and $\m$ becomes
\be%
{\bar  S}_{RN} = \frac{R^3 V_3}{\kappa^2} \frac{1}{T_{RN}} \left[
F(z_+) + \frac{1}{2 z_+^4} + \frac{4 \pi^4 \k^2 \m^2}{3 g^2 R^2
z_+^2} + \frac{8 \pi^4 \k^2 \m^2}{3 g^2 R^2 c z_+^4} \ls e^{-c
z_+^2}  - 1 \rs \right] .
\ee%

Now, we calculate the on-shell action of tcAdS, whose dual theory
lies in the confining phase. Substituting the solutions of
\eq{sol:gauge} and \eq{metric:tcAdS} into \eq{act:Euclidean}, the
on-shell action is obtained as follows
\bea %
S_{tc} = \frac{R^3 V_3}{\kappa^2} \int_0^{\beta'} dt_{\sss{E}}
\int_{\epsilon}^{\infty} dz\, \ e^{- c z^2}  \left( \frac{4}{z^5} -
2 q^{\prime 2} z\right),\nonumber
\eea%
where $\b'$ is the periodicity in the Euclidean time direction. The
integral range over $z$ runs from the UV cut-off $\ep$ to $\infty$
because there is no black hole horizon for tcAdS. After performing
the integration, the on-shell action for tcAdS becomes
\bea%
S_{tc} = \frac{R^3 V_3}{\kappa^2} \beta'\left( - F(\epsilon) -
\frac{1}{c}e^{-c\ep^2}q^{\prime 2} \right),
\eea%
where $F(\infty)=0$ and $H(\infty)=0$ have been used. We renormalize
the on-shell action due to the divergence of $F(\ep)$, as it is done
for RNAdS BH. The circumferences of tcAdS and tAdS at the UV cut-off
should be the same. The relation between the Euclidean time
periodicities is therefore \bea \b_0 = \sqrt{f_{tc} (\epsilon)} \
\b' = \b' + {\cal O} (\ep^6) . \eea By subtracting the tAdS on-shell
action \eq{act:AdS}, the renormalized on-shell action of tcAdS,
${\bar  S}_{tc}$ in the limit as $\ep \to 0$ becomes
\bea  %
\la{act:tc} {\bar  S}_{tc} = S_{tc} - S_{AdS} = - \frac{R^3
V_3}{\kappa^2} \ \b' \ \frac{q'^2}{c}.
\eea%
The grand potential of the dual gauge theory is \be \O_{tc} =
\frac{\bar{S}_{tc}}{\b'} , \ee which is a function of the chemical
potential. So $q^\prime$ should be represented as a function of
$\m$. To determine $q^\prime$ as a function of $\m$, we first assume
a simple relation between $q^\prime$ and $\m$
\bea %
\la{ass:number} q' = \a \m ,
\eea%
where $\a$ is a constant. Using this, the total quark number is
obtained by the thermodynamic relation
\bea %
\la{res:numbertc} N = - \frac{\pa \O_{tc}}{\pa \m} =\frac{R^3
V_3}{\kappa^2} \frac{2
\a q'}{c}.%
\eea %
In \cite{Lee:2009bya}, it was shown that imposing the Neumann
boundary condition instead of the Dirichlet one at the UV cut-off
corresponds to the Legendre transformation from the grand potential
to the free energy and the boundary action $S_b$ is the same as $\m
N \b'$. So, through the calculation of the boundary action we can
determine the unknown parameter $\a$. The boundary action at the UV
cut-off is
\begin{eqnarray}  \la{act:tcbound}
S_b &=& - \frac{1}{g^2} \int_{\partial M} d^4 x \sqrt{G^{(4)}} \, n^{M}\,A_L \,F_{MN}\, g^{LN} \nonumber\\
&=&\frac{V_3 \beta'}{g^2} \,4 \pi^2 R \mu Q ,
\end{eqnarray}
where the unit normal vector is $n^M = \lc 0,0,0,0, - \frac{z}{R}
\sqrt{f_{tc} (z)} \rc$. Comparing \eq{res:numbertc} with
\eq{act:tcbound}, the constant $\a$ is determined as%
\be\la{alpha}\a = 3 c \pi^2 \sqrt{\frac{2 \k^2}{3 R^2 g^2}} . \ee
Substituting \eq{alpha} into \eq{act:tc}, the renormalized action
for tcAdS becomes \be {\bar S}_{tc} = - \frac{R^3 V_3}{\kappa^2} \
\b' \ \frac{\a^2 \m^2}{c} = -\frac{6 c \pi^4 R }{g^2} V_3 \b'
\m^2.\ee
Usually the Hawking-Page transition occurs when the two on-shell
actions for the RNAdS BH and tcAdS are the same. Before describing
it, we first clarify $\b'$. Following \cite{Witten:1998zw}, $\b'$
can be expressed in terms of the Hawking temperature $T_{RN}$ by
matching the time circumferences of these two backgrounds at the UV
cut-off \bea \b' &=& \sqrt{\frac{ f_{RN} (\epsilon) }{f_{tc}
(\epsilon)}} \frac{1}{T_{RN}} \nn &=& \frac{1}{T_{RN}} \ls 1 -
\frac{m}{2} \ep^4 \rs + {\cal O} (\ep^6). \eea From this, the action
difference between the on-shell actions for the RNAdS BH and tcAdS,
which is proportional to the grand potential difference, becomes
\bea %
\D S &\equiv& \bar{S}_{RN} - \bar{S}_{tc} \nn &=& \frac{R^3
V_3}{\kappa^2} \frac{1}{T_{RN}} \left( F(z_+)- \frac{m}{2} \ep^4
F(\epsilon) + H(z_+) -\frac{e^{-c\ep^2}}{c}(q^2-q^{\prime 2})
\right) .\nonumber
\eea%
By expanding the $F(\ep)$  as
\bea%
F (\ep) = - \frac{1}{\ep^4} + {\cal O} \ls \frac{1}{\ep^2} \rs,
\eea%
the on-shell action difference becomes
\bea %
\D S = \frac{R^3 V_3}{\kappa^2} \frac{1}{T_{RN}} \left( \frac{e^{-c
z_+^2}}{z_+^4} \,(c z_+^2 -1) + c^2 {\rm Ei}(-c z_+^2)
 + \frac{1}{2 z_+^4} + \frac{q^2 z_+^2}{2}
+ \frac{e^{-c z_+^2} q^2}{c} -\frac{q^2}{c}+\frac{q'^2}{c}
\right),\nonumber
\eea%
where $z_+$, $q$ and $q^\prime$ are functions of $T_{RN}$ and $\m$
as shown in \eq{rel:zp} and below
\bea %
q=\sqrt{\frac{2}{3}\frac{\kappa^2}{g^2R^2}}\frac{2\pi^2\mu}{z_+^2}
=\sqrt{\frac{2}{3}\frac{N_f}{N_c}}\frac{2\pi^2\mu}{z_+^2},\\
q^\prime=\sqrt{\frac{3}{2}\frac{\kappa^2}{g^2R^2}}2\pi^2\mu
c=\sqrt{\frac{3}{2}\frac{N_f}{N_c}}2\pi^2\mu c.
\eea%
We follow the convention of \cite{Sin:2007ze}
\bea%
\frac{1}{\k^2}=\frac{N_c^2}{4\pi^2R^3},~~\frac{1}{g^2}=\frac{N_cN_f}{4\pi^2R},
\eea%
for the last equalities.

\begin{figure}[h!]
\begin{center}
\subfigure{
\includegraphics[angle=0,width=0.6\textwidth]{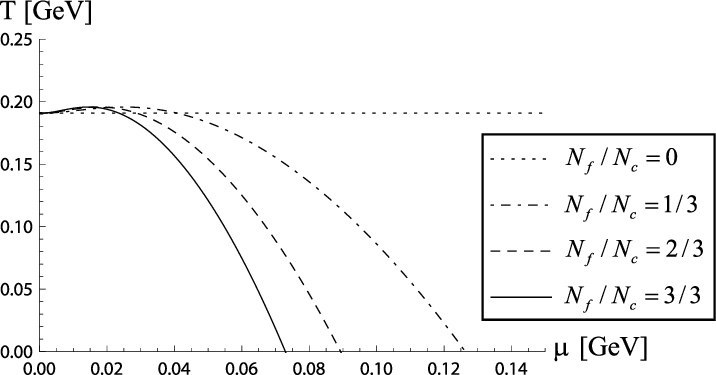}}
\vspace{0cm} \caption{The deconfinement temperature depending on the
chemical potential.} \label{fig:hp}
\end{center}
\end{figure}

The Hawking-Page transition, which corresponds to the
confinement/deconfinement phase transition occurs at $\D S=0$. We
plot the confinement/deconfinement phase diagram for
$N_f/N_c=0,~1/3,~2/3,~3/3$ in Figure \ref{fig:hp}.
%

\section{Meson spectra in the soft wall model}

In this section, we investigate meson masses depending on the quark
number density in the soft wall model. To describe the light mesons,
we concentrate on tcAdS because it corresponds to the confining
phase. Since we will not study the thermodynamic properties of the
meson it is more convenient to use the Lorentzian tcAdS metric \be
\la{met:MtcAdS} ds^2 = \frac{R^2}{z^2} \ls - f_{tc}(z) dt^2 +
\frac{1}{f_{tc}(z)} dz^2 + \d_{ij} dx^i dx^j \rs , \ee where $f_{tc}
(z)$ is
\bea%
f_{tc}(z)=1+q^2z^6,\la{met:fftc}
\eea%
as defined in Section \ref{sec:hp}. The background gauge field
corresponding to the quark number density is \be \la{sol:Mgauge} A_0
= 2 \pi^2 \m - Q z^2 , \ee in the Lorentzian signature.

In the soft wall model \cite{Karch:2006pv}, the gravity action
describing meson spectra of the dual gauge theory is given by
\begin{equation}    \la{act:Mfluc}
S = - \int d^5 x \sqrt{-G} \ e^{-\Phi}\mathrm{Tr} \lb | D \ph |^2 +
m^2 |\ph|^2 + \frac{1}{4 g^2} \ls f^{\sss{(L)}}_{\sss{MN}}
f^{\sss{(L)MN}} + f^{\sss{(R)}}_{\sss{MN}} f^{\sss{(R)MN}} \rs \rb ,
\end{equation}
where $m^2 = -\frac{3}{R^2}$ and the real part of $\ph$ corresponds
to the chiral condensate of the dual gauge theory. Tr means the
trace over the flavor symmetry group indices. The superscripts, $(L)$ and $(R)$, in the
last term imply the left and right parts of $SU(N_f)_L \times
SU(N_f)_R$ flavor symmetry group with $f^{\sss{(L,R)}}_{\sss{MN}} =
\pa_{\sss{M}} a^{\sss{(L,R)}}_{\sss{N}} - \pa_{\sss{N}}
a^{\sss{(L,R)}}_{\sss{M}} - i \lb a^{\sss{(L,R)}}_{\sss{M}},
a^{\sss{(L,R)}}_{\sss{N}} \rb$, where $a_{\sss{M}}$ implies the bulk
fluctuation corresponding to vector or axial-vector mesons. Usually,
there exists non-zero chiral condensate in the confining phase which
breaks the chiral symmetry spontaneously, so we set $\ph$ as
\bea \la{eq:phi}\ph = \frac{v(z)}{2\sqrt{2}}\mathbf{1} e^{i 2\pi^a
T^a} ,\eea
where $v$ is real. $\pi^a$ will be considered as fluctuations
corresponding to pseudoscalar mesons. After turning off all bulk
fluctuations $a_{\sss{M}}$ and $\pi^a$, the equation of motion
describing the chiral condensate is given by \be \la{eq:v} 0 = \pa_z
\ls \frac{f_{tc} }{z^3} \ e^{-\Ph}  \pa_z v \rs + \frac{3}{z^5}
e^{-\Ph} v, \ee
where the gravitational backreaction of $v$ is ignored and $v$ is
considered as a background field describing the chiral condensate
like the original one \ct{Erlich:2005qh}. We set $R=1$. Near the
boundary $z=0$, the asymptotic form of $v$ is \bea \la{eq:scalar}v =
m_q z + \s z^3 +  \cdots , \eea where $m_q$ and $\s$ are the quark
mass and chiral condensate respectively. The existence of the
analytic solution of \eq{eq:v} is not guaranteed so we will use the
numerical solution depending on two initial parameters to
investigate the meson spectra. We investigate the case of $N_f=2$
and $N_c=3$. We set the five-dimensional coupling constant in
\eq{act:Mfluc}, $g^2=\frac{12\pi^2}{N_c}$, following the convention
of \cite{Karch:2006pv}.

\subsection{Vector meson}\label{sec:vec}

We transform the vector fields $a_{\sss{M}}^{\sss{(L)}}$ and
$a_{\sss{M}}^{\sss{(R)}}$ to obtain vector and axial vector fields
\bea v_{\sss{M}} &=& \half \ls a_{\sss{M}}^{\sss{(L)}} +
a_{\sss{M}}^{\sss{(R)}} \rs , \nn a_{\sss{M}} &=& \half \ls
a_{\sss{M}}^{\sss{(L)}} - a_{\sss{M}}^{\sss{(R)}} \rs . \eea
The action describing the vector and axial vector mesons up to the
quadratic order is
\be \la{act:fluc} \D S = - \int d^5x \sqrt{-G} e^{-\Ph} \lb v^2 \ls
\pa_{\sss{M}} \pi - a_{\sss{M}} \rs \ls \pa^{\sss{M}} \pi -
a^{\sss{M}} \rs + \frac{1}{2 g^2} \ls {f^{\sss{(V)}}_{\sss{MN}}}^2 +
{f^{\sss{(A)}}_{\sss{MN}}}^2 \rs \rb ,\ee
where $f^{\sss{(V)}}_{\sss{MN}} = \pa_{\sss{M}} v_{\sss{N}} -
\pa_{\sss{N}} v_{\sss{M}}$ and $f^{\sss{(A)}}_{\sss{MN}} =
\pa_{\sss{M}} a_{\sss{N}} - \pa_{\sss{N}} a_{\sss{M}}$. We study the
model in the gauge where $v_z=0$ and $a_z=0$. Since Lorentz symmetry
is broken we consider the vector field fluctuation in the
$x_i$-direction. The Fourier transform of the vector field can be
defined as
\be \label{eq:vec_fourier} v_i (t,z) = \int \frac{d \o}{2 \pi} e^{-
i \o t} \td{v}_i (z), \ee
with a condition $m_{\sss{V}}^2 = \o^2$. The vector field
$v_{\sss{M}}$ in the action (\ref{act:fluc}) does not couple to the
scalar field $v$. The equation of motion for the vector field $v_i$
under the transformation (\ref{eq:vec_fourier}) is
\be \la{eq:diff} 0 = \pa_z \ls \frac{f_{tc}}{z} e^{- \Ph} \pa_z
\td{v}_i \rs + \frac{m_{\sss{V}}^2}{ z \td{f}_{tc}} e^{- \Ph}
\td{v}_i. \ee
We introduce a dimensionless variable as follows
\bea \la{eq:scale} \td{z} = \sqrt{c} z . \eea
All the other dimensionful parameters get scaled accordingly
\bea f_{tc}(\td{z})=1+\td{q}^2\td{z}^6,~~~\td{q} =
\frac{q}{c^{3/2}},~~~\td{m}_{\sss{V}} =
\frac{m_{\sss{V}}}{\sqrt{c}}. \eea
By the field redefinition $\td{v}_i = e^{- X} \ps_i$ with $X =
\td{z}^2 - \log \frac{f_{tc}}{\td{z}}$, the equation (\ref{eq:diff})
reduces to the Schr\"{o}dinger-like equation
\begin{eqnarray} \la{eq:vec}
0 = \ps_i '' + \ls \frac{X''}{2} - \frac{X'^2}{4}
+\frac{\td{m}_{\sss{V}}^2}{f_{tc}^2} \rs \ps_i.
\end{eqnarray}
Notice that the effective potential $V_{eff} = -  \ls \frac{X''}{2}
- \frac{X'^2}{4} +\frac{\td{m}_{\sss{V}}^2}{f_{tc}^2} \rs$ has the
positive infinite value at two boundaries $z=0$ and $z=\infty$, so
the wave function $\ps_i$ should be zero at these two boundaries.
This implies that the natural boundary conditions for $\ps_i$ are
\bea \label{eq:bdry}\ps_i (0) = \ps_i (\infty) = 0. \eea
In the case of zero quark density ($q=0$, $f_{tc} = 1$), the
effective potential becomes $V_{eff}=\td{z}^2 + \frac{3}{4 \td{z}^2}
- \td{m}_{\sss{V}}^2$ and the equation (\ref{eq:vec}) is exactly
solvable. The solution for the differential equation is
\be \label{eq:laguerre}\ps^{(n+1)}_i = \sqrt{\frac{2n!}{(n+1)!}} \
\td{z}^{3/2} e^{-\td{z}^2 /2} L^1_n (\td{z}^2), \ee
with associated Laguerre polynomials $L^1_n(x)\equiv
\frac{dL_n(x)}{dx}$. This is the $(n+1)$-th excited solution of
\eq{eq:vec} for the quantized mass values $\td{m}_{\sss{V}}^2=
\td{m}_n^2 \equiv 4 (n + 1)$ and corresponds to the $(n+1)$-th
excited vector meson. From this result, we can see the Regge
behavior of the vector meson, $m_n^2 \sim n+1$ , in the zero density
case.

There is no analytic solution for the case of non-zero quark
density. We will investigate the meson spectra numerically. It is
worth noting that the solution (\ref{eq:laguerre}) satisfies the set
of the boundary conditions (\ref{eq:bdry}). It shows that these are
consistent boundary conditions for the numerical analysis.

\begin{figure}[h!]
\begin{center}
\subfigure{
\includegraphics[angle=0,width=0.6\textwidth]{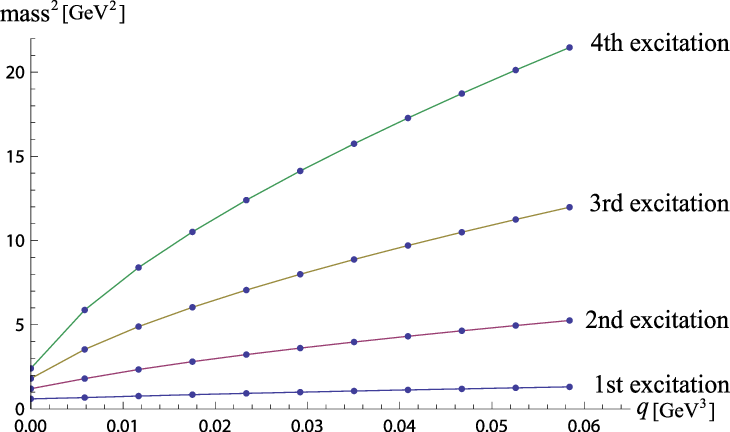}}
\vspace{0cm} \caption{The first four excitations of the vector meson
spectra.} \label{fig:vector}
\end{center}
\end{figure}

In Figure \ref{fig:vector}, we plot vector meson masses for $n+1 =
1,2,3,4$ depending on the quark density. The meson mass increases as
quark density increases. This is qualitatively consistent with the
vector meson spectra in the hard wall model \cite{Jo:2009xr}.

\begin{table}[h!]
\begin{center}
\begin{tabular}{|c||c|c|c|c|}
\hline
$q(\mathrm{GeV}^3)$    & $n=0$& $n=1$& $n=2$ & $n=3$  \\
\hline \hline %
$0.0\times(0.388)^3$    &$0.776$&$1.09743$
         &$1.34407$&$1.55200$\\
\hline%
$0.1\times(0.388)^3$    &$0.82285$&$1.34296$
         &$1.88084$&$2.42465$\\
\hline%
$0.2\times(0.388)^3$    &$0.87629$&$1.53163$
         &$2.21219$&$2.89822$\\
\hline%
$0.3\times(0.388)^3$    &$0.92270$&$1.67681$
         &$2.45737$&$3.24278$\\
\hline%
$0.4\times(0.388)^3$    &$0.96361$&$1.79728$
         &$2.65726$&$3.52160$\\
\hline%
$0.5\times(0.388)^3$    &$1.00035$&$1.90151$
         &$2.82844$&$3.75933$\\
\hline%
$0.6\times(0.388)^3$    &$1.03386$&$1.99413$
         &$2.97948$&$3.96848$\\
\hline%
$0.7\times(0.388)^3$    &$1.06477$&$2.07793$
         &$3.11547$&$4.15639$\\
\hline%
$0.8\times(0.388)^3$    &$1.09274$&$2.15479$
         &$3.23970$&$4.32777$\\
\hline%
$0.9\times(0.388)^3$    &$1.12037$&$2.22599$
         &$3.35445$&$4.48587$\\
\hline%
$1.0\times(0.388)^3$    &$1.14602$&$2.29249$
         &$3.46134$&$4.63300$\\
\hline%
\end{tabular}
\end{center}
\caption{The first four excitations of the vector meson spectra in
GeV. $\sqrt{c}=0.388\mathrm{GeV}$.}
\end{table}

\subsection{Axial vector meson}

\begin{figure}[h!]
\begin{center}
\subfigure{
\includegraphics[angle=0,width=0.6\textwidth]{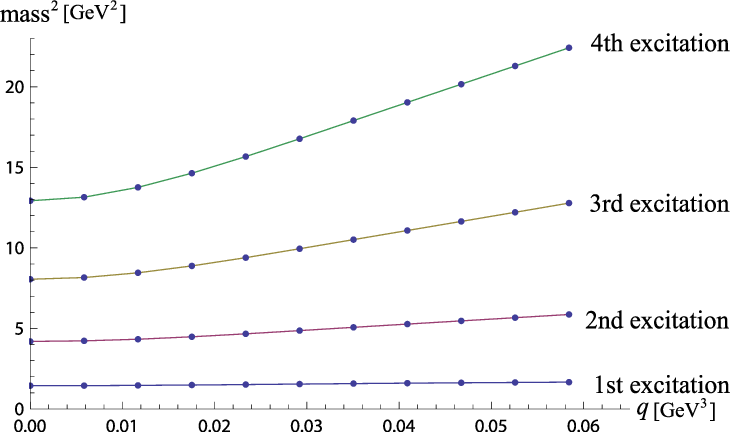}}
\caption{The first four excitations of the axial meson spectra for
$m_q=0.005044\mathrm{GeV}$ and
$\sigma=(0.2619\mathrm{GeV})^3$.}\label{fig:axial}
\end{center}
\end{figure}

We study the axial vector meson spectra. The axial vector can be
decomposed into a transverse component and a longitudinal component
as
\bea a_{\sss{M}}=\bar{a}_{\sss{M}}+\pa_{\sss{M}} \chi. \eea
The $\bar{a}_{\sss{M}}$ corresponds to the axial-vector meson. We
choose the axial gauge, $\bar{a}_{\sss{Z}}=0$. We consider the
fluctuation fixing $\bar{a}_{\sss{0}}=0$ as the Lorentz boost
symmetry is not manifest. As the axial vector couples to the scalar
field $v$ in the action (\ref{act:fluc}) we solve the equation of
motion for the scalar field as well. Under the dimension scale
\eq{eq:scale} the equation \eq{eq:v} becomes
\bea \la{eq:v2}
\pa_{\td{z}}\left(\frac{f_{tc}}{\td{z}^3}e^{-\td{z}^2}\pa_{\td{z}}v\right)
+\frac{3}{\td{z}^5}e^{-\td{z}^2}v=0. \eea
The parameters in the scalar field \eq{eq:scalar} are scaled as
\bea
v=\td{m}_q\td{z}+\td{\sigma}\td{z}^3+\cdots,~~\td{m}_q=\frac{m_q}{\sqrt{c}},~~
\td{\sigma}=\frac{\sigma}{c^{3/2}}. \eea
By the Fourier transformed axial vector
\bea \bar{a}_i(t,z)=\int \frac{d\omega}{2\pi}e^{-i\omega
t}\td{\bar{a}}_i(\omega,z), \eea
the equation of motion 
is obtained as
\bea\la{eq:axial}%
\pa_z\left(\frac{f_{tc}}{z}e^{-\Phi}\pa_z\td{\bar{a}}_i\right)
-\frac{1}{z^3}e^{-\Phi}\left(g^2v^2-\frac{z^2}{f_{tc}}m_{\sss{A}}^2\right)\td{\bar{a}}_i=0.
\eea
By the field redefinition $\td{\bar{a}}_i=e^{Y/2}\zeta_i$ with
$Y=\td{z}^2-\log\frac{f_{tc}}{\td{z}}$, the equation \eq{eq:axial}
can be rewritten as
\bea \la{eq:axial2} {\zeta}''_i+\left(
\frac{Y''}{2}-\frac{Y'^2}{4}-\frac{1}{\td{z}^2f_{tc}}(g^2v^2-\frac{\td{z}^2}{f_{tc}}\td{m}_{\sss{A}}^2)
\right) \zeta_i=0. \eea
The mass spectra from the numerical solution of \eq{eq:v2} and
\eq{eq:axial} for $m_q=0.005044\mathrm{GeV}$ and
$\sigma=(0.2619\mathrm{GeV})^3$ are plotted in Fig \ref{fig:axial}.
The meson mass increases as quark density increases.
\begin{table}[h!]
\begin{center}
\begin{tabular}{|c||c|c|c|c|}
\hline
$q(\mathrm{GeV}^3)$    & $n=0$ & $n=1$& $n=2$ & $n=3$  \\
\hline \hline %
$0.0\times(0.388)^3$    &$1.20195$&$2.04792$
         &$2.83813$&$3.59508$\\
\hline%
$0.1\times(0.388)^3$    &$1.20415$&$2.05663$
         &$2.85662$&$3.62591$\\
\hline%
$0.2\times(0.388)^3$    &$1.21044$&$2.08104$
         &$2.90742$&$3.70920$\\
\hline%
$0.3\times(0.388)^3$    &$1.21996$&$2.11703$
         &$2.98027$&$3.82575$\\
\hline%
$0.4\times(0.388)^3$    &$1.23155$&$2.15995$
         &$3.06494$&$3.95831$\\
\hline%
$0.5\times(0.388)^3$    &$1.24391$&$2.20579$
         &$3.15386$&$4.09546$\\
\hline%
$0.6\times(0.388)^3$    &$1.25581$&$2.25161$
         &$3.24237$&$4.23097$\\
\hline%
$0.7\times(0.388)^3$    &$1.26636$&$2.29591$
         &$3.32850$&$4.36251$\\
\hline%
$0.8\times(0.388)^3$    &$1.27537$&$2.33861$
         &$3.41222$&$4.49003$\\
\hline%
$0.9\times(0.388)^3$    &$1.28327$&$2.38044$
         &$3.49422$&$4.61415$\\
\hline%
$1.0\times(0.388)^3$    &$1.29080$&$2.42216$
         &$3.57504$&$4.73521$\\
\hline%
\end{tabular}
\end{center}
\caption{The first four excitations of the axial meson spectra in
GeV for $m_q=0.005044\mathrm{GeV}$ and
$\sigma=(0.2619\mathrm{GeV})^3$. $\sqrt{c}=0.388\mathrm{GeV}$. }
\end{table}

\section{Discussion}

We have studied the holographic QCD in hadronic medium by using the
soft wall model, where the confinement scale is induced by a
non-dynamical scalar field. We have observed the Hawking-Page
transition between Reissner-Nordstr\"{o}m AdS black hole and thermal
charged AdS and obtained the phase diagram. The patterns of the
plots at high chemical potentials are analogous to the diagrams of
the hard wall model \cite{Lee:2009bya} whereas the patterns of the
plots at low chemical potentials are distinguishable from the
diagrams of the hard wall model. We have studied the vector and
axial vector meson spectra depending on the quark density. In both
cases the meson mass increases as quark density increases. The Regge
behavior is not observed in the presence of the quark density.

We have discussed the model in the backgrounds of RNAdS BH and tcAdS
whose geometries are not modified by the scalar field which induces
the IR cutoff. It would be interesting to study relevant physical
quantities in the geometry whose metric gets deformed by a scalar
field. We will report those results elsewhere.

\section*{Acknowledgements}
We would like to thank S. Sachan for pointing out an error in Figure
1, which appeared in the previous version. This work was supported
by the National Research Foundation of Korea(NRF) grant funded by
the Korea government(MEST) through the Center for Quantum
Spacetime(CQUeST) of Sogang University with grant number
2005-0049409. C. Park was also supported by Basic Science Research
Program through the National Research Foundation of Korea(NRF)
funded by the Ministry of Education, Science and
Technology(2010-0022369).


\end{document}